\documentclass[letterpaper,conference]{IEEEtran}

\usepackage[latin1]{inputenc}
\usepackage[T1]{fontenc}

\usepackage[english]{babel}
\usepackage{color}
\usepackage{amssymb, amsmath}
\usepackage{microtype}
\usepackage{cite}
\usepackage{url}
\usepackage{graphicx}
\usepackage{xcolor}
\usepackage{lipsum}
\usepackage{enumitem}
\usepackage{framed}
\usepackage{textcomp}
\usepackage{subfigure}
\usepackage{balance}
\usepackage{booktabs}
\usepackage{array}
\usepackage[absolute,overlay,showboxes]{textpos}
\usepackage{hyperref}
\usepackage{cleveref}
\hypersetup{
  colorlinks,linkcolor=black,citecolor=black,urlcolor=blue
}
\usepackage{bytefield}
\usepackage{pgfplots}
\usepackage{tikzscale}
\usepackage{tikz}
\usepgfplotslibrary{external}
\usetikzlibrary{positioning,external,calc,shapes,patterns}

\usepackage{pifont}

\usepackage{xspace}

\newcommand{\ie}{\textit{i.e.,}~}
\newcommand{\cf}{\textit{cf.,}~}
\newcommand{\etc}{\textit{etc.}~}
\newcommand{\one}{({\em i})\xspace}
\newcommand{\two}{({\em ii})\xspace}
\newcommand{\three}{({\em iii})\xspace}

\crefname{section}{Section}{Sections}
\crefname{figure}{Figure}{Figures}
\crefname{table}{Table}{Tables}

\makeatletter
\renewcommand{\paragraph}[1]{\vspace*{0.03in}\noindent{\bf #1.}\hspace{0.25ex \@plus1ex \@minus.2ex}}
\makeatother

\makeatletter
\newcommand\footnoteref[1]{\protected@xdef\@thefnmark{\ref{#1}}\@footnotemark}
\makeatother

\definecolor{darkblue}{HTML}{330099}

\title{A Lesson in Scaling 6LoWPAN --\\ Minimal Fragment Forwarding in  Lossy Networks}

\author{%
\IEEEauthorblockN{Martine S. Lenders}
\IEEEauthorblockA{{Freie Universit{\"a}t Berlin}\\
    {m.lenders@fu-berlin.de}}
\and
\IEEEauthorblockN{Thomas C. Schmidt}
\IEEEauthorblockA{{HAW Hamburg}\\
    {t.schmidt@haw-hamburg.de}}
\and
\IEEEauthorblockN{Matthias W{\"a}hlisch}
\IEEEauthorblockA{{Freie Universit{\"a}t Berlin}\\
    {m.waehlisch@fu-berlin.de}}
}

\TPMargin{5pt}

\begin{document}

\maketitle

\begin{textblock}{0.8}(0.1,0.02)
	\noindent
	\footnotesize
	If you cite this paper, please use the LCN reference:
	M. S. Lenders, T. C. Schmidt, M. W\"ahlisch. "A Lesson in Scaling 6LoWPAN - Minimal Fragment Forwarding in Lossy Networks." in \emph{Proc. of IEEE LCN}, 2019.
\end{textblock}

\begin{abstract}
    This paper evaluates two forwarding strategies for fragmented datagrams in the IoT:
    hop-wise reassembly and a minimal approach to direct forwarding of fragments.
    Direct fragment forwarding is challenged by the lack of forwarding information at subsequent fragments in 6LoWPAN and thus requires additional data at nodes.
    We compare the two approaches in extensive experiments evaluating reliability, end-to-end latency, and memory consumption.
    In contrast to previous work and due to our real-world testbed setup, we obtained different results and conclusions. Our findings indicate  that direct fragment forwarding should be deployed with care,  since higher packet transmission rates  on the link layer can significantly reduce its reliability, which in turn can even further reduce end-to-end latency because of highly increased link layer retransmissions.
\end{abstract}


\section{Introduction}\label{sec:intro}

The advent of the Internet of Things (IoT) increased deployment of resource constrained wireless devices in a rapidly growing market.
Always connected sensors and actuators advance business models concerning new products, process innovations, and data.
Wireless operators have already started the  wide-area outreach to the embedded edge, which facilitates operation of IoT gateways in the wild. Foreseeably, 5G technologies appear on the horizon with the promise  of tailored technologies that can host vertical networks towards their users. Such vertical networks, or network slices, will allow public or private bodies and companies to create their own private 5G-based networks on site. This current trend will foster a strong  increase of heterogeneous devices that join the wider Internet, but also a significantly widened range of heterogeneous access networks.

\begin{figure}[t]
    \centering
    \begin{tikzpicture}[x=1, y=1, remember picture, node distance=25]
        \newcommand{\lightbulb}{
            \begin{tikzpicture}[rotate=-45, x=.5, y=.5]
                \draw [fill=white] (0,   0) circle (10);
                \draw [fill=white] (0,  -9) circle [x radius=5, y radius=2.5];
                \draw [fill=white] (0, -10) circle [x radius=5, y radius=2.5];
                \draw [fill=white] (0, -11) circle [x radius=5, y radius=2.5];
                \draw [fill=white] (0, -12) circle [x radius=5, y radius=2.5];
                \draw [fill=white] (0, -13) circle [x radius=5, y radius=2.5];
                \draw [fill=white] (0, -14) circle [x radius=5, y radius=2.5];
            \end{tikzpicture}
        }
        \node (internet) [cloud, draw, cloud puffs=7, cloud puff arc=120, aspect=2, align=center]
            {Internet};
        \node (wpan) [right=of internet, align=center, yshift=6] {
            \begin{tikzpicture}[node distance=5, x=.7, y=.7]
                \draw (-6, 4) -- ++(40, 36) -- ++(40, -36) -- ++(0, -49) -- ++(-80, 0) -- cycle;
                \node (br) [inner sep=0pt] at (0, 0) {
                    \begin{tikzpicture}
                        \draw [rounded corners=.5, fill=white] (-6.5, 0) rectangle ++(1, 12);
                        \draw [rounded corners=3, fill=white] (-10, -4) rectangle ++(20, 8);
                        \draw [rounded corners=2, fill=black] ( -6,  0) circle (.5);
                        \draw [cap=round, line width=1] ( -2,  0) -- (6, 0);
                        \draw [cap=round] (-6.5, 12) ++(25:4) arc (25:-25:4);
                        \draw [cap=round] (-6.5, 12) ++(25:8) arc (25:-25:8);
                    \end{tikzpicture}
                };
                \node (heater) [above right=of br, inner sep=0pt, xshift=-3, yshift=-12] {
                    \begin{tikzpicture}
                        \draw [out=270, in=180, rounded corners=0] (-9, -12) to ++( 1, -2);
                        \draw [out=270, in=  0, rounded corners=0] ( 9, -12) to ++(-1, -2);
                        \draw [out= 90, in=  0, rounded corners=0] ( 9,  12) to ++(-1,  2);
                        \draw [out= 90, in=180, rounded corners=0] (-9,  12) to ++( 1,  2);
                        \draw                                      ( 9, -12) -- ( 9,  12);
                        \draw                                      (-9, -12) -- (-9,  12);
                        \draw [out=  0, in=180, rounded corners=0] (-8, -14) to
                            ++(2, 2) to ++(2, -2) to ++(2, 2) to ++(2, -2) to
                            ++(2, 2) to ++(2, -2) to ++(2, 2) to ++(2, -2);
                        \draw [out=  0, in=180, rounded corners=0] (-8,  14) to
                            ++(2, -2) to ++(2, 2) to ++(2, -2) to ++(2, 2) to
                            ++(2, -2) to ++(2, 2) to ++(2, -2) to ++(2, 2);
                        \draw [rounded corners=1] (-5, -10) rectangle ++(-2, 20);
                        \draw [rounded corners=1] (-1, -10) rectangle ++(-2, 20);
                        \draw [rounded corners=1] ( 1, -10) rectangle ++( 2, 20);
                        \draw [rounded corners=1] ( 5, -10) rectangle ++( 2, 20);
                    \end{tikzpicture}
                };
                \node (thermometer) [below right=of heater, inner sep=0pt, xshift=2, yshift=9] {
                    \begin{tikzpicture}
                        \draw [fill=white] (0,0) circle (3);
                        \draw [fill=white] (0,20) circle (1.5);
                        \draw [fill=white, draw=none, rounded corners=0] (-1.5,0) rectangle ++(3, 20);
                        \draw (-1.5, 20) -- ++(0, -17.5);
                        \draw ( 1.5, 20) -- ++(0, -17.5);
                        \draw [fill=white] (0,20) circle (0.5);
                        \draw [fill=white] (0,0) circle (2);
                        \draw [fill=white, draw=none, rounded corners=0] (-0.5,0) rectangle ++(1, 20);
                        \draw (-0.5, 20) -- ++(0, -18.25);
                        \draw ( 0.5, 20) -- ++(0, -18.25);
                    \end{tikzpicture}
                };
                \node (lightbulb1) [below=of br, inner sep=0pt, yshift=-5] {\lightbulb};
                \node (switches) [below=of thermometer, inner sep=0pt] {
                    \begin{tikzpicture}
                        \draw [rounded corners=2.25] (-7, -1) rectangle ++(14, -6);
                        \draw [rounded corners=2.25] (-7,  1) rectangle ++(14,  6);
                        \draw (-4, -4) circle (3);
                        \draw (-4, -4) circle (1.5);
                        \draw ( 4,  4) circle (3);
                        \draw ( 4,  4) circle (1.5);
                    \end{tikzpicture}
                };
                \draw (br) -- (heater) -- (thermometer);
                \draw (br) -- (lightbulb1);
                \draw (lightbulb1) -- (switches);
                \draw (lightbulb1) -- (thermometer);
            \end{tikzpicture}
        };
        \node (plc) [left=of internet, align=center, draw, rounded corners=5] {
            \begin{tikzpicture}[node distance=20, x=.7, y=.7]
                \node (modem) [inner sep=0pt, anchor=west] at (0, 0) {
                    \begin{tikzpicture}
                        \draw [rounded corners=3, fill=white] (-10, -4) rectangle ++(20, 8);
                        \draw [rounded corners=2, fill=black] ( -6,  0) circle (.5);
                        \draw [cap=round, line width=1] ( -2,  0) -- (6, 0);
                    \end{tikzpicture}
                };
                \node (powerpole1) [inner sep=0pt, left=of modem, anchor=north,
                                    rounded corners=0] {
                    \begin{tikzpicture}
                        \node (powerpole1-connect-in1) at (5, 0) {};
                        \node (powerpole1-connect-out1) at (5, 0) {};
                        \node (powerpole1-connect-out2) at (0, 0) {};
                        \node (powerpole1-connect-out3) at (-5, 0) {};
                        \draw [cap=round, line width=1] (5, 0) -- ++(0, -2) -- ++(-10, 0) -- (-5, 0);
                        \draw [cap=round, line width=1] (0, 0) -- ++(0, -30);
                    \end{tikzpicture}
                };
                \draw [bend left] (modem) to (powerpole1-connect-in1);
                \node (powerpole2) [inner sep=0pt, left=of powerpole1, anchor=north, yshift=0,
                                    rounded corners=0] {
                    \begin{tikzpicture}
                        \node (powerpole2-connect-in1) at (5, 0) {};
                        \node (powerpole2-connect-out1) at (5, 0) {};
                        \node (powerpole2-connect-in2) at (0, 0) {};
                        \node (powerpole2-connect-in3) at (-5, 0) {};
                        \draw [cap=round, line width=1] (5, 0) -- ++(0, -2) -- ++(-10, 0) -- (-5, 0);
                        \draw [cap=round, line width=1] (0, 0) -- ++(0, -30);
                    \end{tikzpicture}
                };
                \draw [bend left] (powerpole1-connect-out1) to (powerpole2-connect-in1);
                \draw [bend left] (powerpole1-connect-out2) to (powerpole2-connect-in2);
                \draw [bend left] (powerpole1-connect-out3) to (powerpole2-connect-in3);
                \node (house) [inner sep=0pt, right=of powerpole2, anchor=north, xshift=20, yshift=3,
                               rounded corners=0] {
                    \begin{tikzpicture}
                        \node (house-in) at (0, 0) {};
                        \draw (0, 0) -- ++(10, 10) -- ++(10, -10) -- ++(0, -15) -- ++(-20, 0) -- cycle;
                        \draw (3, -15) -- ++(0, 13) -- ++(6, 0) -- ++(0, -13);
                        \draw (12, -7) -- ++(0, 5) -- ++(5, 0) -- ++(0, -5) -- cycle;
                    \end{tikzpicture}
                };
                \draw [bend right] (powerpole2-connect-out1) to (house-in);
            \end{tikzpicture}
        };
        \draw (modem) -- (internet) -- (br);
        \node [below=of wpan, anchor=north, yshift=28] {IoT (127~bytes)};
        \node [below=of internet, anchor=north, yshift=15.5] {(1500~bytes)};
        \node [below=of plc, anchor=north, yshift=25] {PLC (400~bytes)};
    \end{tikzpicture}
    \caption{A typical scenario where datagram fragmentation is needed
    because of different maximum packet sizes.}
    \label{fig:intro:iot-scenario}
\end{figure}
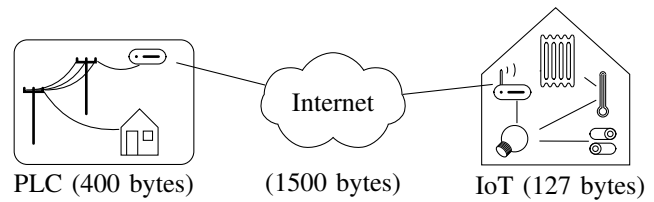

Besides the wireless IoT, other access technologies such as Power-line Communication (PLC) gather deployment, while offering a wide range of packet sizes \cite{draft-ietf-6lo-plc}.
These different technologies introduce a wide variety of maximum packet sizes in the link layer as visualized in \cref{fig:intro:iot-scenario}.
On the network layer, nodes predominantly speak IPv6 \cite{RFC-8200} with a mandatory transparent Maximum Transmission Unit (MTU) size of at least 1280 bytes.
Hence, fragmentation is necessary in order to communicate using these link layer technologies.

Some of these links---e.g. IEEE 802.15.4 \cite{IEEE-802.15.4-15}---only support a very limited number of bytes.
For efficiency, information required to forward a packet cannot be encoded in every fragment but is only present in the first~\cite{RFC-4944}. This is in contrast to transparent fragmentation such as in the IP protocols.
Since many IoT networks form meshes, however, forwarding of packets is needed, and there are two concepts for forwarding fragmented datagrams.
First, reassembly is performed at every hop (\emph{hop-wise reassembly}), followed by re-fragmentation when forwarded on another constrained link.
This is the simplest solution, due to the forwarding information only being stored in the first fragment.
Second, individual fragments are forwarded (\emph{fragment forwarding}) by recording the forwarding information required from the first fragment on all participating nodes.
This recorded information then can be used to forward all subsequent fragments to the next hop~\cite[Section~2.5.2]{sb-6wei-09},~\cite{draft-ietf-6lo-minimal-fragment}.

Both approaches---hop-wise reassembly and fragment forwarding---have advantages and disadvantages.
While direct forwarding can lead to lower latency, it also sends more packets on average over time, leading to a higher load on the medium.
Hop-wise reassembly on the other hand is part of common network stacks, which can be a benefit on more constrained nodes, where program memory is scarce \cite{RFC-7228}.

In this paper, we comparatively assess the performance and resource consumption of \emph{hop-wise reassembly} and direct \emph{fragment forwarding} over a thin  IEEE 802.15.4 MAC layer. Our findings are ambivalent and reveal two sides of the coin. Depending on the MAC layer and packet frequency, hop-wise reassembly may perform much better than the prospective optimization introduced with fragment forwarding. Conversely, MAC layers with a slow coordinative function like IEEE 802.15.4e can profit from fragment forwarding.
As part of this work, we also provide an independent implementation of fragment forwarding, which we showcase to allow deeper insights into our evaluation~results.

The remainder of this paper is structured as follows. In \cref{sec:background}, the background  of 6LoWPAN fragmentation and forwarding is recapitulated along with related work.
\cref{sec:impl} describes our implementation for fragment forwarding, with which we obtain  the results presented in \cref{sec:eval}. We discuss our findings in \cref{sec:discussion}, and close with a conclusion and an outlook in \cref{sec:conclusion}.

\section{Problem Statement and Related Work}\label{sec:background}
\newcommand{\rb}[2]{%
    \node [draw] at #1 [anchor=west, right, fill=white, opacity=.9, inner sep=1pt] {
        \begin{tikzpicture}[y=8]
            \node [anchor=east, left] at (-0.5, -1.075) {\scriptsize $\mathbf{id}(i)$};
            \draw [fill=white,draw=none] (-0.5, -.7) rectangle ++(1,-.75);
            \draw [fill=darkblue!80,draw=none] (-0.5,  -.7) rectangle ++(#2,-.75);
            \draw (-0.5,-.7) rectangle ++(8,-.75);
        \end{tikzpicture}
    }
}
\newcommand{\frag}[3]{
    \draw [->] ($#1 + #2$) --
        node [anchor=center, fill=white, inner sep=0.5pt] {$f_{i#3}$}
        ++(2,1);
}
\newcommand{\frags}[2]{%
    \frag{#1}{(0, 0)}{1};
    \frag{#1}{($(22, 0) + (0, #2)$)}{2};
    \frag{#1}{($(54, 0) + (0, #2)$)}{n};
    \node [anchor=mid] at ($#1 + (38.5, .5) + (0, #2)$) {$\dots$};
}
\newcommand{\reass}[1]{%
    \frags{#1}{0};
    \rb{($( 3, 1) + #1$)}{2};
    \rb{($(25, 1) + #1$)}{4};
    \rb{($(57, 1) + #1$)}{8};
}
\newcommand{\timeline}[3]{%
    \node [anchor=east] at #1 {$#2$};
    \draw [line width=.75] #1 -- ++(#3, 0);
}
\newcommand{\vrbadd}[3]{%
    \node [draw] at #3 [anchor=south west, fill=white,
                        opacity=.9, inner sep=1pt] {
        \scriptsize $i \mapsto (\mathbf{id}(i), #2, \mathbf{t}_{#2}(i))$%
    };
    \draw [<-, dashed, opacity=.75] #3 -- #1;
}
\newcommand{\vrblookup}[3]{%
    \node [draw] at #3 [anchor=south west, fill=white,
                        opacity=.9, inner sep=1pt] {
        \scriptsize $\mathbf{id}(i) \mapsto (#2, \mathbf{t}_{#2}(i))$%
    };
    \draw [<-, dashed, opacity=.75] #3 -- #1;
}

The IETF specified the 6LoWPAN protocol~\cite{RFC-4944} to allow for transmissions of IPv6 packets over IEEE 802.15.4~\cite{IEEE-802.15.4-15} networks---a widely used link layer technology in the IoT.
While IPv6 requires a Maximum Transmission Unit~(MTU) of at least 1280 bytes \cite{RFC-8200}, IEEE 802.15.4 is only able to handle link layer packets of up to 127 bytes---\emph{including} the link layer header---which in the worst case only leaves 33 bytes for application data~\cite{kkt-siotsp-14}.
To enable IPv6 communication in such a restrictive environment, 6LoWPAN provides both header compression \cite{RFC-6282, RFC-7400} and datagram fragmentation~\cite{RFC-4944}.
The latter is the focus of this paper.

For completeness we note that the concept of 6LoWPAN (or more generally \emph{6Lo}) is not limited to IEEE 802.15.4, but also can be used in other link-layer technologies such as PLC~\cite{draft-ietf-6lo-plc}.

\subsection{Basic Fragmentation and Reassembly in 6LoWPAN}\label{sec:background:6lo}
In 6LoWPAN, datagram fragmentation implements the following common approach:
Before sending a datagram to the underlying link layer, the network layer checks whether the data exceeds the maximum payload length (commonly referred to as SDU, Service Data Unit) of the link layer.
If the data size complies with the SDU, a single datagram is sent without any modification.
If the data size does not comply with the SDU, a datagram is divided into multiple fragments such that the content of each fragment matches the SDU.
Each fragment includes a fragment header containing information to assemble the datagram~\cite{RFC-4944}:

The fragmentation header of the first fragment contains an (uncompressed) datagram size in bytes as an 11-bit number and a 16-bit datagram tag to identify the fragment on the link.
All subsequent fragments carry in addition to the header fields of the first fragment header an offset to this fragment in units of 8 bytes, see \cref{fig:background:6lo-frag}.
Consequently, all payloads in a fragment must be of a length that is a multiple of 8.

The receiver identifies multiple fragments that belong to the same datagram by comparing three values: \one the link layer source and destination addresses, \two the datagram size, and \three the datagram tag.
Then, the receiver network stack stores all fragments of an incoming datagram in the \emph{reassembly buffer} for up to 10~seconds.
These identifying parameters to assign fragments to a datagram~$i$ we will refer to by $\mathbf{id}(i)$ in the following.

A brief back-of-the-envelope calculation shows that a node needs to allocate at least \textbf{1302 bytes} of memory \textbf{per reassembly buffer entry} to reassemble a fragmented datagram:
\begin{itemize}
    \item At most \textbf{8 bytes} per address, plus \textbf{1 byte} per address to store their length as IEEE 802.15.4 supports both 64-bit EUI-64s and a 16-bit short addresses as addressing format,
    \item \textbf{2 bytes} for the datagram size,
    \item \textbf{2 bytes} for the datagram tag, and
    \item \textbf{1280 bytes} for the maximum expected size of an IPv6 datagram.
\end{itemize}

1302~bytes are significant memory requirements on constrained devices, which typically offer memory within the range of several kilobytes~\cite{RFC-7228}.
Especially in a multihop network---a common deployment scenario in the IoT---it becomes challenging to provide enough resources to store a sufficient number of reassembly buffer entries.
In \cref{sec:impl:6lo}, we show how to save memory in a concrete implementation.

\begin{figure}[t]
    \centering
    \scalebox{.9}{
    \begin{tikzpicture}[x=25, y=25]
        \node at (0,0.5) [anchor=east, align=right] {Datagram (DG)};
        \draw [draw, pattern=north east lines, pattern color=darkblue!80] (0,0) rectangle ++(6.5,1)
            node [midway, above, yshift=20, anchor=mid]
            {$i$};
        \draw [fill=darkblue!40] (0,0) rectangle ++(1,1);
        \node at (.5,0.5) [anchor=mid] {\scriptsize DG hdr};

        \node at (-1,-1.5) [anchor=east, left, align=right] {Fragmented};
        \draw [draw, pattern=north east lines, pattern color=darkblue!80] (-1,-2) rectangle ++(2.5,1)
            node [midway, below, yshift=-20, anchor=mid]
            {$f_{i1}$};
        \draw [fill=darkblue!20] (-1,-2) rectangle ++(1,1);
        \draw [draw, line width=.1pt] (-1,-1.5) -- ++(1,0);
        \draw [fill=darkblue!40] (0,-2) rectangle ++(1,1);
        \node at (-.5,-1.25) [anchor=mid] {\scriptsize DG size};
        \node at (-.5,-1.75) [anchor=mid] {\scriptsize DG tag};
        \node at (.5,-1.5) [anchor=mid] {\scriptsize DG hdr};

        \draw [draw, pattern=north east lines, pattern color=darkblue!80] (1.8,-2) rectangle ++(2.5,1)
            node [midway, below, yshift=-20, anchor=mid]
            {$f_{i2}$};
        \draw [fill=darkblue!20] (1.8,-2) rectangle ++(1.3,1);
        \draw [draw, line width=.1pt] (1.8,-1.5) -- ++(1,0);
        \draw [draw, line width=.1pt] (2.8,-2) -- ++(0,1);
        \node at (2.3,-1.25) [anchor=mid] {\scriptsize DG size};
        \node at (2.3,-1.75) [anchor=mid] {\scriptsize DG tag};
        \node at (2.95,-1.5) [anchor=mid, rotate=90] {\scriptsize offset};
        \draw [draw=none] (4.3, -1) -- ++(1,-1) node [midway, anchor=mid] {$\dots$};

        \draw [draw, pattern=north east lines, pattern color=darkblue!80] (5.3,-2) rectangle ++(2.5,1)
            node [midway, below, yshift=-20, anchor=mid]
            {$f_{in}$};
        \draw [fill=darkblue!20] (5.3,-2) rectangle ++(1.3,1);
        \draw [draw, line width=.1pt] (5.3,-1.5) -- ++(1,0);
        \draw [draw, line width=.1pt] (6.3,-2) -- ++(0,1);
        \node at (5.8,-1.25) [anchor=mid] {\scriptsize DG size};
        \node at (5.8,-1.75) [anchor=mid] {\scriptsize DG tag};
        \node at (6.45,-1.5) [anchor=mid, rotate=90] {\scriptsize offset};

        \draw [dashed] (0,0) -- (0,-1);
        \draw [dashed] (1.5,0) -- (1.5,-1);
        \draw [dashed] (1.5,0) -- (3.1,-1);
        \draw [dashed] (2.8,0) -- (4.3,-1);
        \draw [draw=none] (3.55, -0.5) -- (5.9,-0.5) node [anchor=mid, midway, align=center, yshift=1.5pt] {$\dots$};
        \draw [dashed] (5.2,0) -- (6.6,-1);
        \draw [dashed] (6.5,0) -- (7.8,-1);
    \end{tikzpicture}
    }
    \caption{Fragmentation in 6LoWPAN.}
    \label{fig:background:6lo-frag}
\end{figure}
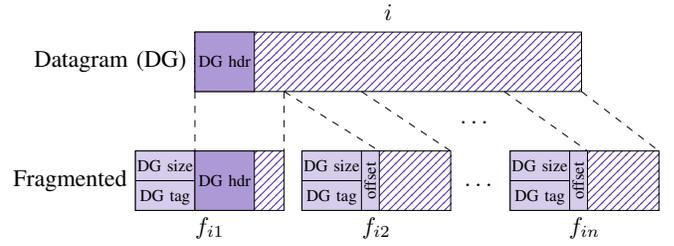

\begin{figure}[b!]
    \centering
    \begin{tikzpicture}[node/.style={circle,draw,anchor=mid,
                                     minimum size=20,
                        node distance=.3}]
        \node (r1) {$\dots$};
        \node [node, left=of r1] (f) {$f$};
        \node [node, fill=darkblue!40, left=of f] (e) {$e$};
        \node [node, above left=of e] (b) {$b$};
        \node [node, left=of b] (a) {$a$};
        \node [left=.3 of a] (r2) {$\dots$};
        \node [node, below left=of e] (d) {$d$};
        \node [node, left=of d] (c) {$c$};
        \node [left=.3 of c] (r3) {$\dots$};

        \draw [->] (r2) edge (a) (a) edge (b) (b) edge (e);
        \draw [->] (r3) edge (c) (c) edge (d) (d) edge (e);
        \draw [->] (e) edge (f) (f) edge (r1);
    \end{tikzpicture}
    \caption{$e$ represents a typical bottleneck for HWR.}
    \label{fig:background:hwr-bottleneck}
\end{figure}
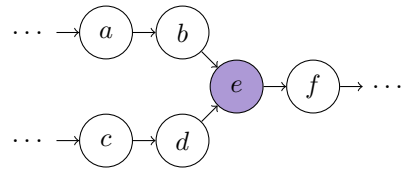
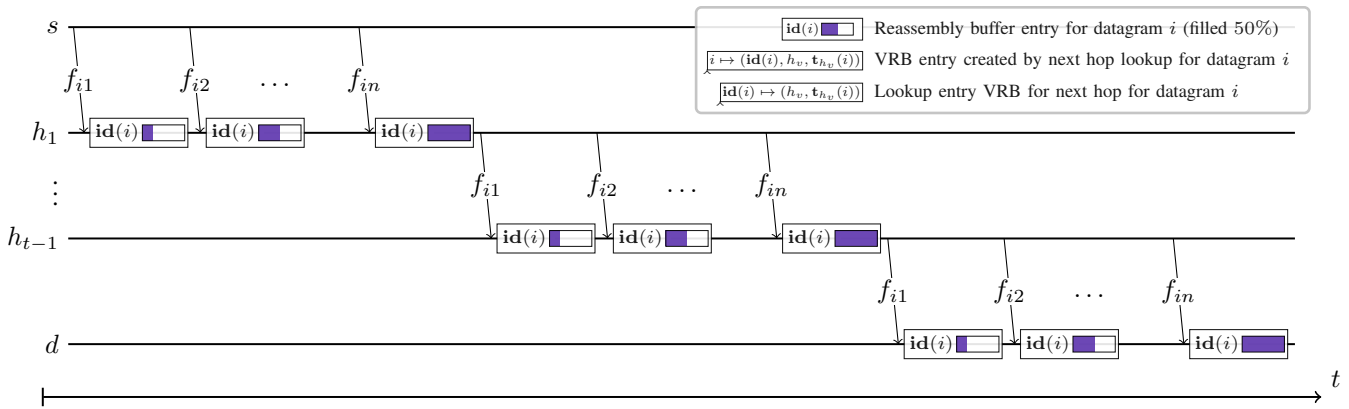
\begin{figure*}[h]
    \centering
    \begin{tikzpicture}[x=2,y=-40]
        \draw [|->, line width=.75] (-5, 3.5) -- ($(232, 3.5) + (5, 0)$) node [anchor=south west] {$t$};
        \node [anchor=east] at (0, 1.5) {$\vdots$};
        \timeline{(0, 0)}{s}      {232};
        \timeline{(0, 1)}{h_1}    {232};
        \timeline{(0, 2)}{h_{t-1}}{232};
        \timeline{(0, 3)}{d}      {232};

        \reass{(  1, 0)};
        \reass{( 78, 1)};
        \reass{(155, 2)};

        \node at (235, -0.2) [anchor=north east, align=right, fill=white, opacity=.9, inner sep=2pt,
                              rounded corners=2pt, line width=1pt, draw=lightgray] {
            \begin{tikzpicture}
                \node at (0, 0) [anchor=east, left, align=right] {
                    \scalebox{.75}{%
                        \begin{tikzpicture}[rounded corners=0]
                            \rb{(0,0)}{4};
                        \end{tikzpicture}%
                    }%
                };
                \node at (0, 0) [anchor=west, right, align=left] {\scriptsize Reassembly buffer entry for datagram~$i$ (filled $50\%$)};

                \node at (0, .3) [anchor=east, left, align=right] {
                    \scalebox{.75}{%
                        \begin{tikzpicture}[rounded corners=0, rotate=0]
                            \vrbadd{(0,0)}{h_v}{(0, 0)};
                        \end{tikzpicture}%
                    }%
                };

                \node at (0, .3) [anchor=west, right, align=left] {\scriptsize VRB entry created by next hop lookup for datagram~$i$};
                \node at (0, .6) [anchor=east, left, align=right] {
                    \scalebox{.75}{%
                        \begin{tikzpicture}[rounded corners=0, rotate=0]
                            \vrblookup{(0,0)}{h_v}{(0, 0)};
                        \end{tikzpicture}%
                    }%
                };
                \node at (0, .6) [anchor=west, right, align=left] {\scriptsize Lookup entry VRB for next hop for datagram~$i$};
            \end{tikzpicture}
        };
    \end{tikzpicture}
    \caption{Hop-wise Reassembly (HWR) of a datagram~$i$ ($s$: source; $h_1, \dots, h_{t-1}$: intermediate hops; $d$: destination; $\mathbf{t}_{x}(i)$: datagram tag to next hop $x$).}
    \label{fig:background:hwr}
\end{figure*}

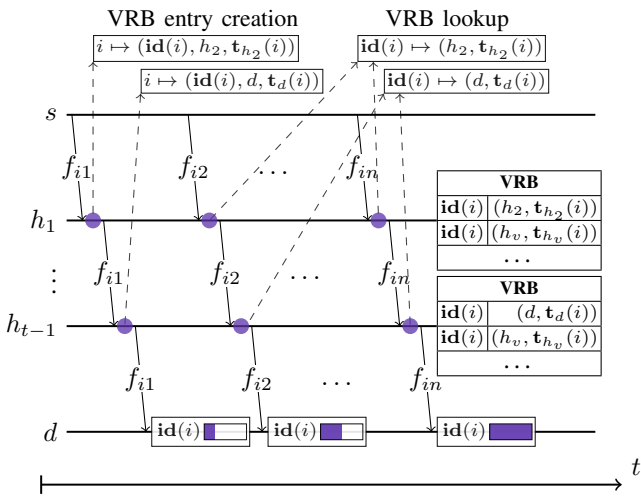
\begin{figure}[t]
    \centering
    \begin{tikzpicture}[x=2,y=-40]
        \draw [|->, line width=.75] (-5, 3.5) -- ($(100, 3.5) + (5, 0)$) node [anchor=south west] {$t$};
        \node [anchor=east] at (0, 1.5) {$\vdots$};
        \timeline{(0, 0)}{s}      {100};
        \timeline{(0, 1)}{h_1}    {70};
        \timeline{(0, 2)}{h_{t-1}}{70};
        \timeline{(0, 3)}{d}      {100};

        \frags{( 1, 0)}{0};
        \frags{( 7, 1)}{0};
        \reass{(13, 2)};

        \vrblookup{(27,  1)}{h_2}{(55, -.5)};
        \draw [<-, dashed, opacity=.75] (58, -.5) -- (59, 1);
        \vrblookup{(34,  2)}{d}{(60, -.2)};
        \draw [<-, dashed, opacity=.75] (63, -.2) -- (65, 2);
        \vrbadd{( 5,  1)}{h_2}{(5, -.5)};
        \vrbadd{(11,  2)}{d}{(14, -.2)};
        \draw [fill=darkblue!80, opacity=.75, draw=none] (5, 1) circle (3pt);
        \draw [fill=darkblue!80, opacity=.75, draw=none] (11, 2) circle (3pt);
        \draw [fill=darkblue!80, opacity=.75, draw=none] (27, 1) circle (3pt);
        \draw [fill=darkblue!80, opacity=.75, draw=none] (33, 2) circle (3pt);
        \draw [fill=darkblue!80, opacity=.75, draw=none] (59, 1) circle (3pt);
        \draw [fill=darkblue!80, opacity=.75, draw=none] (65, 2) circle (3pt);
        \node at (70, 1) [anchor=west, right, xshift=-7pt] {
                \def\arraystretch{0.75}
                \begin{tabular}{|>{\hspace{-5pt}\scriptsize}r<{\hspace{-5pt}}|>{\hspace{-5pt}\scriptsize}r<{\hspace{-5pt}}|}
                    \hline
                    \multicolumn{2}{|c|}{\scriptsize\textbf{VRB}} \\\hline
                    $\mathbf{id}(i)$ & $(h_2,\mathbf{t}_{h_2}(i))$ \\\hline
                    $\mathbf{id}(i)$ & $(h_v,\mathbf{t}_{h_v}(i))$ \\\hline
                    \multicolumn{2}{|c|}{\dots} \\\hline
                \end{tabular}
            };
        \node at (70, 2) [anchor=west, right, xshift=-7pt] {
                \def\arraystretch{0.75}
                \begin{tabular}{|>{\hspace{-5pt}\scriptsize}r<{\hspace{-5pt}}|>{\hspace{-5pt}\scriptsize}r<{\hspace{-5pt}}|}
                    \hline
                    \multicolumn{2}{|c|}{\scriptsize\textbf{VRB}} \\\hline
                    $\mathbf{id}(i)$ & $(d,\mathbf{t}_{d}(i))$ \\\hline
                    $\mathbf{id}(i)$ & $(h_v,\mathbf{t}_{h_v}(i))$ \\\hline
                    \multicolumn{2}{|c|}{\dots} \\\hline
                \end{tabular}
            };
        \node [anchor=south, align=center] at (25.75, -.7) {\small VRB entry creation};
        \node [anchor=south, align=center] at (72, -.7) {\small VRB lookup};
    \end{tikzpicture}
    \caption{Fragment Forwarding (FF) of a datagram~$i$ using the Virtual Reassembly Buffer (VRB). Notations comply with \cref{fig:background:hwr}.}
    \label{fig:background:mff}
\end{figure}

\subsection{Fragment Forwarding for Low-power Lossy Networks}\label{sec:background:minfwd}

The destination address in the IPv6 header guides forwarding.
In 6LoWPAN fragmentation, however, the IPv6 header is only present in the first fragment.
To enable intermediate nodes in a multihop network to forward fragments without this context information, two solutions are proposed: hop-wise reassembly and direct fragment forwarding.

The naive approach to handle fragmented datagrams in a multihop network is \emph{hop-wise reassembly~(HWR)}~\cite{RFC-4944,RFC-8200}.
In HWR, each intermediate hop between source and destination assembles and re-fragments the original datagram completely.
This leads to three drawbacks.
First, each intermediate hop needs to provide enough memory resources to store all fragments in the reassembly buffer (see \cref{fig:background:hwr}).
Second, the memory requirements are unbalanced between nodes in the network.
Considering highly connected nodes (see node~$e$ in \cref{fig:background:hwr-bottleneck}), these nodes need to cope with the reassembly load of all their downstream nodes.
Third, datagram delivery time is bound by the time needed to receive all fragments of the datagram.
Papadopoulos~\emph{et al.}~\cite{pttm-rpfri-18} underscored these problems in more detail.

\emph{Fragment forwarding~(FF)}~\cite{draft-ietf-6lo-minimal-fragment} tackles the drawbacks of HWR by leveraging a \emph{virtual} reassembly buffer \emph{(VRB)}~\cite{draft-ietf-lwig-6lowpan-virtual-reassembly}, see \cref{fig:background:mff}.
In contrast to a reassembly buffer, a VRB only stores references to link the subsequent fragments to the first fragment such that intermediate nodes can determine the next hop.
In detail, the VRB is applied as follows.
Each entry represents the source and destination addresses, the datagram size, the datagram tag ($\mathbf{id}(i)$, \cf \cref{sec:background:6lo}), the next hop link layer address $h_v$, and the outgoing datagram tag $\mathbf{t}(i)$.
This has two implications.
First, an intermediate node can ensure that datagram tags are unique between a node and its neighbors.
Second, all fragments belonging to the same datagram will travel the same path.

\subsection{Additional Related Work}\label{sec:background:related-work}

Other approaches that use similar concepts as FF mainly focus on datagram prioritization~\cite{w-fsf6id-17, wrtt-roft6-13}.
In addition to FF, the 6lo working group of the IETF is also working on a forwarding mechanism that includes \emph{selective fragment recovery}~\cite{draft-ietf-6lo-fragment-recovery}.
Selective fragment forwarding is effectively a complete new fragmentation protocol introducing new header types for 6LoWPAN.
As it allows for recovery of lost fragments and provides congestion control mechanisms it could help to mitigate the congestion problems we observed in our experiments.
Exploring its advantages in more detail will be part of our future work.

Similar to this, Chowdhury \emph{et al.}~\cite{cicrky-rmr6-09} proposed a standard compliant NACK-based approach for selective fragment recovery.
Since those NACKs, however, are associated with $\mathbf{id}(i)$, this mechanism only allows for hop-wise recovery and does not cover the whole end-to-end path when using FF.

The work most closely related~\cite{tmw-6ff-19} to our study uses the 6TiSCH simulator~\cite{egvw-s6n-18} to analyze the performance of FF.
The authors show that FF is a promising option in IEEE 802.15.4e (TSCH).
As part of our experiments, we revisit the results of these experiments in a real-world setting and find that the abstraction in the simulation of~\cite{tmw-6ff-19} leads to misleading conclusions.

Awwad \emph{et al.}~\cite{anahi-6refr-18} also compared FF to HWR in a simulator and conducted experiments in a testbed.
They used a topology consisting of 4~nodes in a line.
This setup ignores challenging bottlenecks, which occur in real deployments (see \cref{sec:background:minfwd}).
Furthermore, they only compared their proprietary solution of fragment forwarding with HWR in the testbed evaluation.
In contrast to this, we evaluate standard compliant protocols in a complex testbed setup.

In our work, we did not consider the frame delivery mode for link-layer meshes of 6LoWPAN~\cite{RFC-4944}---commonly known as \emph{mesh-under}~\cite[Section 1.2]{sb-6wei-09}---because it is known that such a solution falls behind HWR~\cite{cicrky-rmr6-09}.

\section{Implementation}\label{sec:impl}
A thorough experimental evaluation of protocols requires sound software implementations.
For the sake of comparison, the protocols under investigation should be analyzed on the same system.
Unfortunately, there is no software basis available which assembles all required components for constrained devices.
In this paper, therefore, we extend RIOT~\cite{bhgws-rotoi-13,bghkl-rosos-18}, a common IoT operating system.
By selecting an open source platform and making our software publicly available we enable reproducible research~\cite{swgsc-terrc-17,acmrep}.
Based on our extensions, we gain detailed insights into system and network performance.

In the remainder of this section, we present design, implementation, and configuration choices to better understand the subsequent evaluation.

\subsection{System Details on 6LoWPAN}\label{sec:impl:6lo}

RIOT provides a stable 6LoWPAN implementation as part of its default network stack, GNRC~\cite{plwhb-ownsr-15,lkhpg-cwemr-18}.
Instead of statically allocating packet space for each reassembly buffer, it uses the preconfigurable packet allocation arena of GNRC, called \texttt{gnrc\_pktbuf}, to dynamically allocate packet buffer space of varying length within it.
This allows for high resource efficiency and flexibility.
By storing the major part of the IPv6 datagram (1280 bytes) only in the packet buffer, the 6LoWPAN stack requires \textbf{22 bytes} (plus some additional bytes for management), instead of allocating the complete 1302 bytes (\cf Section~\ref{sec:background:6lo}).

To provide low delays and high throughput, the fragmentation is done asynchronously.
For this purpose, the reference to the datagram that needs to be fragmented is stored in a fragmentation buffer.
The data of the datagram resides in \texttt{gnrc\_pktbuf}.
In addition to the datagram, the fragmentation buffer also contains meta-information needed for fragmentation, including the original datagram size and its tag.

\subsection{Fragment Forwarding}
We extend 6LoWPAN in GNRC to support direct \emph{fragment forwarding}.\footnote{\url{https://github.com/RIOT-OS/RIOT/pull/11068}}
One crucial implementation choice relates to the creation of the first fragment.
The first fragment may include the compression header~\cite{RFC-6282}, which may change size during network traversal as compression contexts such as link-layer addresses change.
Because of that, the compression may be less or more effective depending on header updates made by intermediate forwarders.
In the worst case, the packet becomes less compressed, leading to additional fragmentation.
To tackle this problem, we apply a well-known approach by keeping the first fragment as minimal as possible~\cite{draft-ietf-lwig-6lowpan-virtual-reassembly}, \ie the original sender includes only the fragment and compression headers and pushes the payload to the subsequent fragment.
It is worth noting that this approach does not increase the overall number of fragments compared to a naive approach that minimizes the size of the last fragment.
In fact, it will reduce the likely creation of additional fragments.

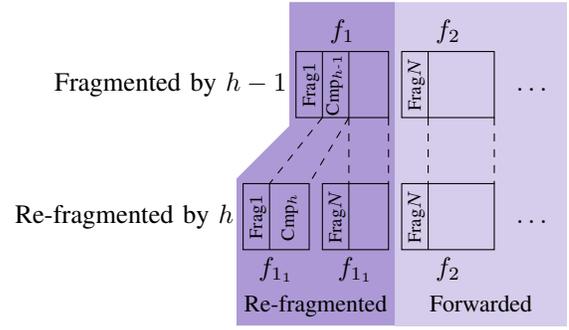
\begin{figure}[t]
    \centering
    \begin{tikzpicture}[x=10, y=25]
        \draw [fill=darkblue!40, draw=none] (3.75, 1.75) --
                                   ++(-4, 0) --
                                   ++(0, -1.875) --
                                   ++(-2, -0.8) --
                                   ++(0, -2.225) --
                                   ++(6, 0) -- cycle;
        \draw [fill=darkblue!20, draw=none] (3.75, 1.75) --
                                        ++(6.5, 0) --
                                        ++(0, -4.9) --
                                        ++(-6.5, 0) -- cycle;

        \node at (0,0.5) [anchor=mid, left, align=right] {Fragmented by $h-1$};
        \draw [draw] (0,0) rectangle ++(3.5,1)
            node [midway, above, yshift=20, anchor=mid]
            {$f_{1}$};
        \node at (0.5,0.5) [anchor=mid, rotate=90] {\scriptsize Frag$1$};
        \draw [draw] (1,0) -- ++(0,1);
        \node at (1.4,0.5) [anchor=mid, rotate=90] {\scriptsize Cmp$_{h \text{-} 1}$};
        \draw [draw] (2,0) -- ++(0,1);

        \draw [draw] (4,0) rectangle ++(3.5,1)
            node [midway, above, yshift=20, anchor=mid]
            {$f_{2}$};
        \node at (4.5,0.5) [anchor=mid, rotate=90] {\scriptsize Frag$N$};
        \draw [draw] (5,0) -- ++(0,1);

        \node at (9,0.5) [anchor=mid] {$\dots$};

        \node at (-2,-1.5) [anchor=mid, left, align=right] {Re-fragmented by $h$};
        \draw [draw] (-2,-2) rectangle ++(2.5,1)
            node [midway, below, yshift=-20, anchor=mid]
            {$f_{1_1}$};
        \node at (-1.5,-1.5) [anchor=mid, rotate=90] {\scriptsize Frag$1$};
        \draw [draw] (-1,-2) -- ++(0,1);
        \node at (-0.3,-1.5) [anchor=mid, rotate=90] {\scriptsize Cmp$_{h}$};

        \draw [draw] (1,-2) rectangle ++(2.5,1)
            node [midway, below, yshift=-20, anchor=mid]
            {$f_{1_1}$};
        \node at (1.5,-1.5) [anchor=mid, rotate=90] {\scriptsize Frag$N$};
        \draw [draw] (2,-2) -- ++(0,1);

        \draw [draw] (4,-2) rectangle ++(3.5,1)
            node [midway, below, yshift=-20, anchor=mid]
            {$f_{2}$};
        \node at (4.5,-1.5) [anchor=mid, rotate=90] {\scriptsize Frag$N$};
        \draw [draw] (5,-2) -- ++(0,1);

        \node at (9,-1.5) [anchor=mid] {$\dots$};

        \draw [dashed] (1,0) -- (-1,-1);
        \draw [dashed] (2,0) -- (0.5,-1);
        \draw [dashed] (2,0) -- (2,-1);
        \draw [dashed] (3.5,0) -- (3.5,-1);
        \draw [dashed] (5,0) -- (5,-1);
        \draw [dashed] (7.5,0) -- (7.5,-1);
        \node at (7,-2.85) [anchor=mid] {\small Forwarded};
        \node at (0.75,-2.85) [anchor=mid] {\small Re-fragmented};
    \end{tikzpicture}
    \caption{Compression header (Cmp) handling for fragment forwarding in the RIOT GNRC.}
    \label{fig:impl:iphc-ff}
\end{figure}

We support this mechanism not only on the original sender but also on intermediate forwarders for the case that the original sender did not provide enough space for the expanding compression header, see Figure~\ref{fig:impl:iphc-ff}.
This is possible, as all subsequent fragments also contain an offset, which indicates fragmentation relating to the first fragment.
Furthermore, it simplifies the implementation greatly, which in turn saves ROM.
Since the fragmentation buffer is used for this, its default size of 1 needs to be increased so that the node is able to handle multiple datagrams---forwarded datagrams and datagrams sent by the node itself---at the same time.

To keep the implementation simple, we only forward fragments when the first fragment is received in order, otherwise we reassemble the packet completely.
This can be considered a fall-back to hop-wise reassembly.

\subsection{MAC Layer}\label{sec:impl:l2}
In its default configuration, GNRC only provides a very slim MAC layer that benefits from radio drivers that support CSMA/CA, link layer retransmissions, and acknowledgement handling by default.
Special care has to be taken for hardware platforms that use ``blocking wait on send'' whenever the device is in a busy state.
When deploying fragment forwarding, this may cause race conditions within the internal state machine of the device \cite{microchip:at86rf231} because of the faster interchange of simultaneous sending and receiving events.
To solve this problem, we provide a simple mechanism to queue packets whenever the device signals that it is in a busy state. 
As soon as the device becomes available again (and not later than 5~ms), the MAC layer tries to send the packet from the top of the queue again.

\section{Evaluation}\label{sec:eval}

Our evaluations are performed in a real-world testbed using class-2 IoT nodes~\cite{RFC-7228} and real 802.15.4 radio communication. One important aspect of the experiment design is the underlying network topology, which we consider by selecting specific nodes from the testbed. We want to assure that \one the network is widespread enough and not too crowded, but also that \two it contains multiple bottlenecks as described in \cref{sec:background:minfwd} to stress hop-wise reassembly.

Our goal is to carefully explore the behavior of the competing fragmentation schemes and along this line to reproduce  simulation results of \cite{tmw-6ff-19}. From many previous experiences we know that simulation---although an important tool for network analysis---often produces misleading results in the complex and surprising world of low-power wireless communication.

\subsection{Setup}\label{sec:eval:setup}

\begin{figure}[t]
    \subfigure[Logical topography.]{\includegraphics[width=.48\linewidth]{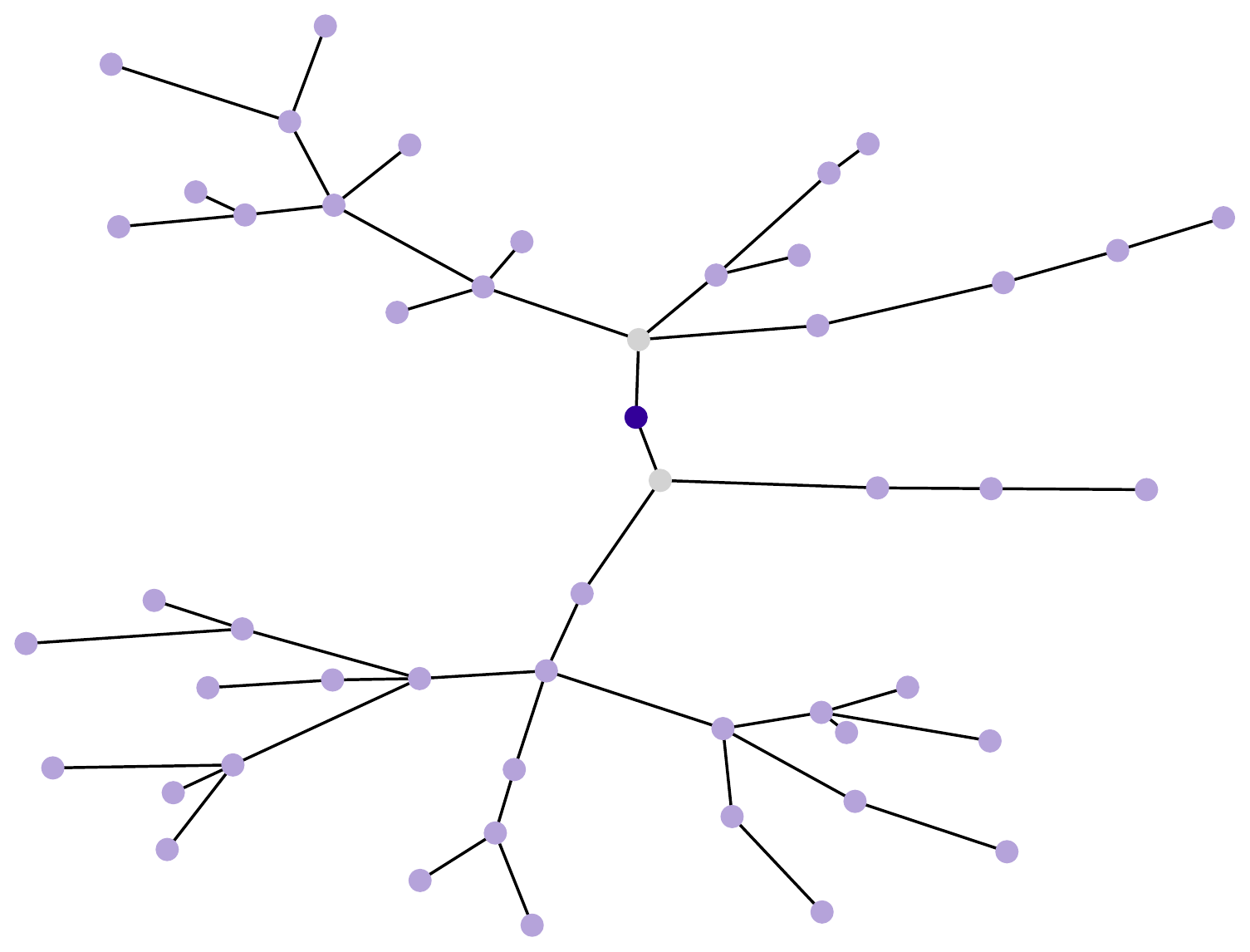} \label{fig:eval:network:topo}}
    \subfigure[Geographical topography.]{\includegraphics[width=.48\linewidth]{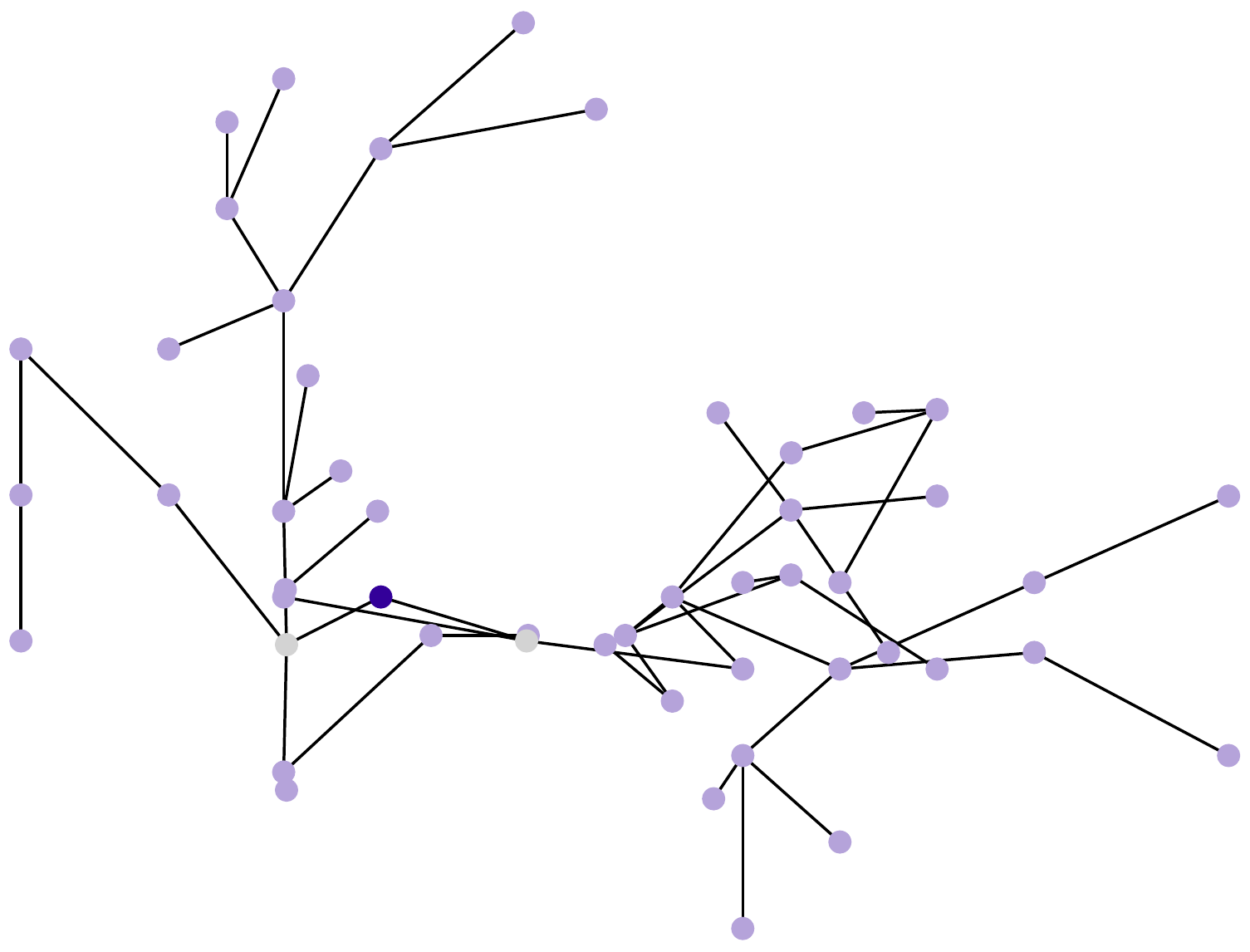} \label{fig:eval:network:geo}}
    \caption{Topography of the selected testbed network (dark-blue: sink, light-blue: source nodes).}
    \label{fig:eval:network}
\end{figure}

\paragraph{Experiment Testbed and Node Selection}
We deploy our experiments on the FIT IoT-LAB testbed and use 50~nodes of the Lille site.
These are constrained IoT devices with Cortex-M3 MCUs, 64 kB of RAM, 512 kB of ROM (STM32F103REY), and IEEE 802.15.4 radios (Atmel AT86RF231).
The radio chip provides the basic MAC layer features such as CSMA/CA, link layer retransmissions, and acknowledgements.

The Lille site features a challenging multihop network.
Nodes are not only distributed in a dedicated room in a grid but also located in multiple offices spread over different floors.
The site therefore provides a realistic scenario for different types of heterogeneous deployment.
However, a careful selection of nodes is necessary to control side effects that may negatively affect our observations.

To select nodes for our experiment, we first measure basic properties of the testbed.
By correlating the geographic distance and the packet delivery ratio (PDR) between two nodes, we found that two hops should be in range of 6.6~m or less.
This ensures that the PDR is at least 97.5\%, which we argue is acceptable.
Lower PDRs do not contribute to a better understanding of the problem space in this paper.
The network is then constructed by a breadth-first search over all available nodes of the testbed site, starting at the sink $s$.\footnote{We select node 55 as the sink as it is located centrally between the more crowded nodes in the dedicated room and the more sparse nodes in the office space at the Lille site.
This ensures that a balanced set of both network deployment scenarios is included.}
To prevent a bias towards specific nodes, our network construction algorithm works as follows.
\begin{enumerate}
    \item Collect all neighbors within the range of 2.2~m and 6.6~m as potential node candidates in set $N$. This selection expands the network as much as possible under our PDR requirement.
    \item Get a randomized, uniformly distributed sample $M$ of 1 to 3 members in $N$; $s$ always selects 2 neighbors.
    \item Add $M$ to the network, and continue for each member of $M$ until 49~nodes are found.
\end{enumerate}

The selection of 1 to 3 downstream neighbors per node assures the inclusion of reassembly bottlenecks into the network, as described in \cref{sec:background:minfwd}.

\begin{table}[t]
    \centering
    \caption{Fragments / UDP payload sizes mapping.}\label{tab:eval:frag-size}
    \begin{tabular}{rrrr}
    \toprule
        Fragments \# & UDP Payload & \multicolumn{2}{c}{Mean Reassembly Time} \\
                                     \cmidrule(lr){3-4}
        & & HWR & FF \\
    \midrule
         1 &   16 bytes & \multicolumn{2}{c}{(no reassembly)} \\
         2 &   80 bytes &  4.3 ms &  5.8 ms \\
         3 &  176 bytes & 10.8 ms & 13.7 ms \\
         4 &  272 bytes & 17.4 ms & 19.6 ms \\
         5 &  368 bytes & 23.9 ms & 26.2 ms \\
         6 &  464 bytes & 32.4 ms & 33.5 ms \\
         7 &  560 bytes & 37.3 ms & 39.1 ms \\
         8 &  656 bytes & 45.2 ms & 47.5 ms \\
         9 &  752 bytes & 52.4 ms & 54.5 ms \\
        10 &  848 bytes & 57.4 ms & 60.6 ms \\
        11 &  944 bytes & 64.1 ms & 67.1 ms \\
        12 & 1040 bytes & 71.3 ms & 73.1 ms \\
        13 & 1136 bytes & 78.7 ms & 80.7 ms \\
        14 & 1232 bytes & 85.2 ms & 88.0 ms \\
    \bottomrule
    \end{tabular}
\end{table}

After constructing the network, we used the same set of nodes in all of our experiments to ensure comparability.
The resulting logical and geographical topologies are visualized in \cref{fig:eval:network}.
Multiple paths have the same length.
The longest path consists of 6~hops.

\paragraph{Communication Setup}
We configured all routes based on the breadth-first search.
Except for the sink and its neighbors, we configured all other nodes as data senders to ensure the need for forwarding.

All source nodes start sending UDP packets---using the same payload---to the sink in a uniformly distributed interval between 5~s and 15~s.
The experiment ends after each source has sent 100~packets.
In contrast to the reference simulation~\cite{tmw-6ff-19}, we select a smaller interval to allow for a significant number of runs.
Slower sending rates would lead to unfeasible durations in our real-world experiments.
It is worth noting that our decision is made carefully:
We conducted one experiment with exactly the same run times as described in the related work.
The results are consistent with our experiments that adapt the improved parameter setting.
The same is the case for smaller network sizes.

HWR and FF implement different fragmentation strategies (\cf \cref{sec:impl:6lo}).
Consequently, the original UDP payload may lead to differently sized fragments resulting in varying overheads for the reassembly processes.
To allow for the fair comparison of both approaches, we need to align the baseline depending on the UDP payload size.
\cref{tab:eval:frag-size} shows the best results based on our empirical validation.
We use these payload sizes in our subsequent experiments.

To evaluate the performance, our experiments measure the same metrics as the simulation~\cite{tmw-6ff-19}.
This includes reliability, specifically the PDR, and the latency between the UDP sockets of source and sink.
In addition, we also assess system complexity in terms of memory.

\begin{figure*}[t]
    \subfigure[Maximum packet buffer usage per fragment multiplicity.]{%
        \includegraphics{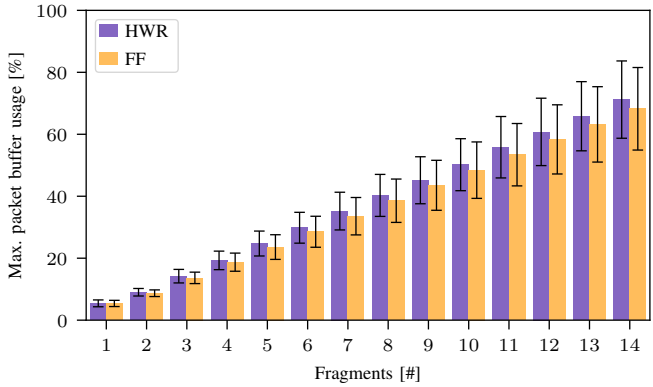}%
        \label{fig:eval:pktbuf:frag}%
    }
    \hfill
    \subfigure[Maximum packet buffer usage versus reassembly buffer exhaustion for FF. The saturation of a hexagon indicates higher multiplicity of these events coinciding in its area.]{%
        \includegraphics{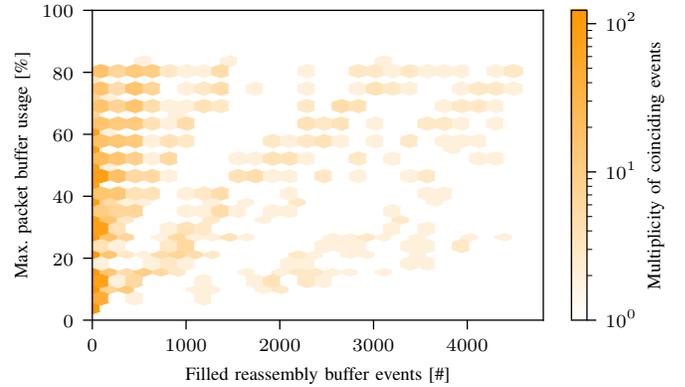}%
        \label{fig:eval:pktbuf:vs-rbuf}%
    }
    \caption{Analysis of the packet buffer utilization.}
    \label{fig:eval:pktbuf}
\end{figure*}

\paragraph{Software Parameterization} \label{sec:impl:param}
RIOT offers a variety of compile-time configuration parameters to adapt to use cases.
In most of the experiments, we can use default configurations.
For the following reasons, however, we have to change some default values:
\one The default configurations assume rather small networks.
This conflicts with efficient forwarding in large-scale mesh networks, such as our testbed.
\two We want to compare our results with related work that analyzed some aspects in simulation~\cite{tmw-6ff-19}.
We document the changes of default values in the Appendix.

In contrast to the parameters in \cite{tmw-6ff-19}, the default size of the virtual reassembly buffer in GNRC is 16~bytes.
Since this only prefers direct forwarding, we do not need to adapt its size.
Furthermore, we have to increase the size of the common reassembly buffer of the sink.
Without this adaptation the reliability decreases significantly, even for the smallest number of fragments.

\subsection{Result 1: Memory Consumption}\label{sec:eval:mem}
\begin{table}[b]
    \centering
    \caption{Memory sizes [bytes] for source nodes.}
    \label{tab:eval:mem}
    \begin{tabular}{rrrrr}
        \toprule
        Module            & \multicolumn{2}{c}{HWR}                           & \multicolumn{2}{c}{FF} \\
                                 \cmidrule(lr){2-3}                                  \cmidrule(lr){4-5}
                          & \multicolumn{1}{c}{ROM} & \multicolumn{1}{c}{RAM} & \multicolumn{1}{c}{ROM} & \multicolumn{1}{c}{RAM} \\
        \midrule
        6LoWPAN            &                    5950 &                    6124 &                    6472 &                    4284 \\
        VRB               &                      n/a &                     n/a &                     316 &                     768 \\
        Forwarding        &                      n/a &                     n/a &                     544 &                       0 \\
        \cmidrule{1-5}
        Sum               &                    5950 &                    6124 &                    7332 &                    5052 \\
        \bottomrule
    \end{tabular}
\end{table}

\cref{tab:eval:mem} shows both ROM and RAM usage of the 6LoWPAN layer at the source node for both forwarding approaches.
When compiling the software we use \texttt{arm-none-eabi-gcc} v7.3.1 with \texttt{-Os} optimization (size-optimal) for ARM Cortex-M3 and the compile-time parameters we line out in \cref{sec:impl:param}. We use the \texttt{size} tool to extract the relevant module information.
To make memory measurements compatible, we set the reassembly buffer size to the same value as the VRB size (16) for HWR.
The anticipated memory advantage does indeed exist, even with the GNRC strategy to not allocate 1280 bytes IPv6 MTU for every reassembly buffer entry but using the central packet buffer instead (\cf \cref{sec:impl}).

FF adds a small amount of RAM to keep the meta-data required for refragmentation in the asynchronous GNRC fragmentation buffer.
More ROM is also needed for the possible refragmentation of the first fragment.
The majority of the $\approx$~500~bytes of additional ROM for FF in 6LoWPAN is explained by the overhead required to distinguish whether packets need to be handled by a VRB entry creation or put into the regular reassembly buffer.

\cref{fig:eval:pktbuf} presents our analysis of the actual utilization of the 6144 bytes packet buffer.
For FF the packet buffer is used just a little less than for HWR.
This can be seen in \cref{fig:eval:pktbuf:frag}, which plots the maximum packet buffer utilization during the runtime of each experiment.
The high packet buffer usage for FF is mostly caused by the fallback to regular reassembly  as we describe in more detail in \cref{sec:eval:rel}.

A clear correlation between events where a full reassembly buffer coincides with a high (or low) maximum usage of the packet buffer can be seen in \cref{fig:eval:pktbuf:vs-rbuf}. This plot visualizes events taken from all nodes during three runs of the experiment.
More saturated hexagons indicate higher multiplicities of events in this area.
The occurrence of many full reassembly buffers tends to lead to high maximum packet buffers. Those coinciding events, however, are less likely in general.
The observed clusters are in line with the hop distances from the nodes on which the coinciding events happen to the sink.

\subsection{Result 2: Reliability and Latency}\label{sec:eval:rel}
\begin{figure*}[h]
    \subfigure[Packet delivery ratio.]{
        \includegraphics{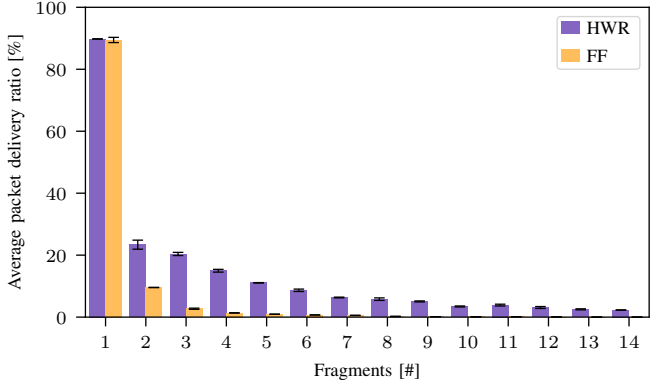}
        \label{fig:eval:vanilla:pdr}}
    \hfill
    \subfigure[Source-to-sink latency (socket-to-socket) by hop distance.]{
        \includegraphics{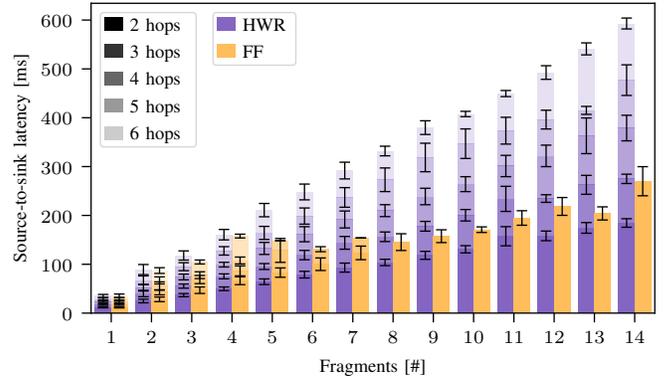}
        \label{fig:eval:vanilla:lat}}
    \subfigure[Link layer retransmissions per node. Lines depict average values.]{
        \includegraphics{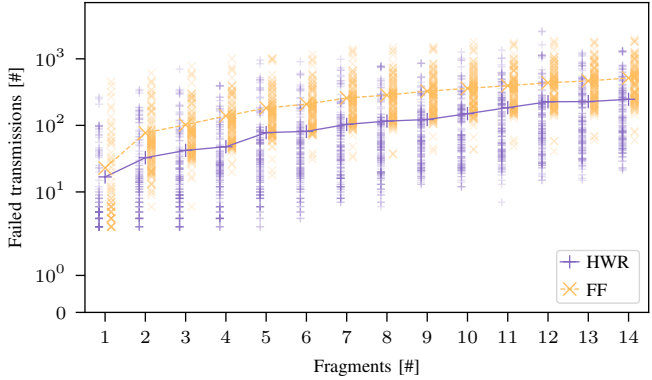}
        \label{fig:eval:vanilla:l2_retrans}}
    \hfill
    \subfigure[Filled reassembly buffer events per node. Lines depict average values.]{
        \includegraphics{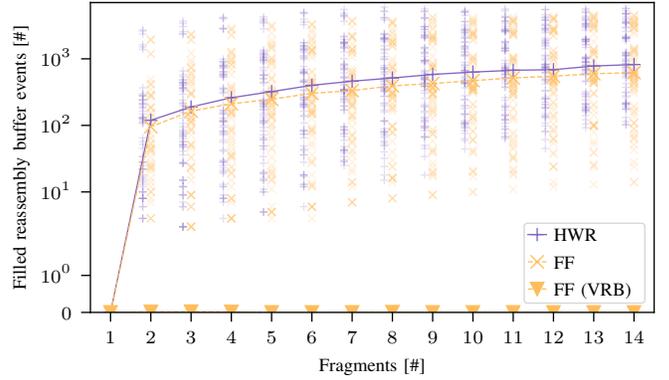}
        \label{fig:eval:vanilla:rbuf_full}}
    \caption{Measurement results for 100 packets every [5,10] s per node (3 runs). }
    \label{fig:eval:vanilla}
\end{figure*}

\cref{fig:eval:vanilla:pdr,fig:eval:vanilla:lat} displays our results from measuring reliability and latency.
Strikingly, FF  admits poor reliability, which is in contrast to previous results  \cite{tmw-6ff-19}.
Even for a small number of fragments, FF achieves less than half the PDR of HWR. Values then quickly approach zero with increasing number of fragments. HWR, though also performing poorly, manages to deliver at least some packets to the more distant nodes.

The latencies we measured for FF are also significantly higher than in the previous simulation work. HWR is expected to operate slower because each node needs to reassemble the entire frame prior to forwarding to the next hop.

\begin{figure}[t]
    \centering
    \begin{tikzpicture}[x=5, y=-7]
        \node [anchor=south] at  (0,0) {$a$};
        \node [anchor=south] at (20,0) {$b$};
        \node [anchor=south] at (40,0) {$c$};

        \draw [fill=darkblue!40,draw=none, rounded corners=2]  (-1,1) rectangle ++(2,4);
        \draw [fill=darkblue!40,draw=none, rounded corners=2]  (-1,6) rectangle ++(2,4);
        \draw [fill=darkblue!40,draw=none, rounded corners=2]  (19,2) rectangle ++(2,3);
        \draw [fill=darkblue!40,draw=none, rounded corners=2]  (19,6) rectangle ++(2,4);
        \draw [fill=darkblue!40,draw=none, rounded corners=2] (39,7) rectangle ++(2,3);

        \draw  (0, 0) -- ++(0, 10.5);
        \draw (20, 0) -- ++(0, 10.5);
        \draw (40, 0) -- ++(0, 10.5);

        \node [anchor=north, yshift=6, xshift=0.075] at  (0,10.5) {\footnotesize $\vdots$};
        \node [anchor=north, yshift=6, xshift=0.075] at (20,10.5) {\footnotesize $\vdots$};
        \node [anchor=north, yshift=6, xshift=0.075] at (40,10.5) {\footnotesize $\vdots$};

        \draw [->] (0, 1) --
            node [sloped, anchor=center, fill=white, inner sep=0pt]
                {\scriptsize $f_{1}[\mathrm{llr}=0]$}
            ++(20,1);
        \draw [->] (20, 3.5) --
            node [sloped, anchor=center, fill=white, inner sep=0pt]
                {\scriptsize $f_{1}[\mathrm{ack}]$}
            ++(-20,1);

        \draw [->] (20, 6) --
            node [sloped, anchor=center, fill=white, inner sep=0pt]
                {\scriptsize $f_{1}[\mathrm{llr}=0]$}
            ++(20,1);
        \draw [->] (40, 8.5) --
            node [sloped, anchor=center, fill=white, inner sep=0pt]
                {\scriptsize $f_{1}[\mathrm{ack}]$}
            ++(-20,1);

        \draw [->] (0, 6) --
            node [sloped, anchor=center, fill=white, inner sep=0pt]
                {\scriptsize $f_{2}[\mathrm{llr}=0]$}
            ++(20,1);
        \draw [->] (0, 7) --
            node [sloped, anchor=center, fill=white, inner sep=0pt]
                {\scriptsize $f_{2}[\mathrm{llr}=1]$}
            ++(20,1);
        \draw [->] (0, 8) --
            node [sloped, anchor=center, fill=white, inner sep=0pt]
                {\scriptsize $f_{2}[\mathrm{llr}=2]$}
            ++(20,1);
        \node at (40, 0) [anchor=north east, xshift=-3pt, align=right, fill=white,
                          opacity=.9, inner sep=2pt, rounded corners=2pt, line width=.5pt,
                          draw=lightgray] {
            \begin{tikzpicture}
                \node at (0, 0) [anchor=east, left, align=right, inner sep=0pt] {
                    \begin{tikzpicture}
                        \draw [fill=darkblue!40,draw=none, rounded corners=2]  (-1,1) rectangle ++(2,1);
                    \end{tikzpicture}
                };
                \node at (0, 0) [anchor=west, right, align=left, inner sep=0pt] {\scriptsize Device busy};
                \node at (0, 1.3) [anchor=east, left, align=right, inner sep=0pt] {
                    \scriptsize $f_{i}[\mathrm{llr}=x]$
                };
                \node at (0, 1.3) [anchor=west, right, align=left, inner sep=0pt] {\scriptsize $x$-th L2 retrans.};
                \node at (0, 2.3) [anchor=west, right, align=left, inner sep=0pt] {\scriptsize of frag. $i$};
                \node at (0, 3.6) [anchor=east, left, align=right, inner sep=0pt] {
                    \scriptsize $f_{i}[\mathrm{ack}]$
                };
                \node at (0, 3.6) [anchor=west, right, align=left, inner sep=0pt] {\scriptsize ACK for frag. $i$};
            \end{tikzpicture}
        };
    \end{tikzpicture}
    \caption{Example of L2 retransmissions caused by the device being busy, resulting in missing ACKs.}
    \label{fig:eval:busy-state}
\end{figure}
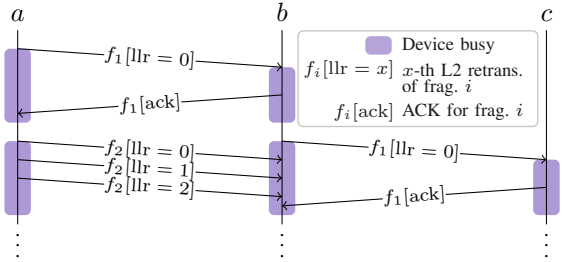

To explore the underlying reasons for the poor performance of FF, we analyse the radio transmission and media occupancy.
\cref{fig:eval:vanilla:l2_retrans} plots the number of link layer retransmissions that occurred for each node within the network over three experiment runs as a scatter plot with a logarithmically scaled y-axis.
The line plot within the scatter plot represents the means of the respective data set.

In our experiments, we  see significantly more link layer retransmissions per node with FF than with HWR.
This is caused by much faster  send and receive triggers on the device due to immediate fragment forwarding, which  increases collisions and packet loss.
Moreover, this results in straining the single buffer of a device, which far more often needs to discard unacknowledged incoming packets while it is busy with either sending or receiving a different packet.
This invokes link layer retransmissions and eventually contributes to packet loss.
An example for these occurrences is illustrated in \cref{fig:eval:busy-state}.
We are able to confirm with local measurements on a sister device of the nodes' radio (AT86RF233 \cite{microchip:at86rf233}) that the device can remain busy for up to 4 ms utilizing a logic analyzer.
With HWR the link is more relaxed due to the time it takes to reassemble and re-fragment a packet again, which leaves both the device and the medium non-stressed.

We can also see that packets are lost with FF when the respective reassembly buffers are full.
In \cref{fig:eval:vanilla:rbuf_full}, we plot these occurrences of reassembly buffer exhausts for each node over 3 experiment runs analog to \cref{fig:eval:vanilla:l2_retrans}.
The reassembly buffer with FF is only about $23\%$ less often filled than with HWR.
This hints at frequent transmissions that lose the first fragment and cause the FF implementation to fall back to normal reassembly, since the first fragment is missing or is received out of order (\cf \cref{sec:impl:6lo}).
Indeed, in 50-60\% of the cases, we observe that the reassembly buffer expires because the first fragment of a datagram is missing.
These transmission failures fill the reassembly buffer up with incomplete datagrams, especially when more fragments are lost.
In this scenario a datagram never actually takes the full~10~s of reassembly timeout to reassemble at each hop (the plain source-to-sink latency  is 600~ms at most). Hence it is unlikely that different strategies---for instance re-fragmenting partly reassembled datagrams and forwarding the rest as soon as the first fragment comes in---would noticeably increase reliability.
Still, this process might save space in the reassembly buffer:
When the first fragment just arrives after a subsequent fragment the reassembly buffer does not require the full space of the datagram.

To further verify that reliability problems are not caused by our implementation, we repeat the experiments with a modified version of FF.
Our modification makes FF simulate the behavior of HWR by putting the fragments to forward in a VRB-associated queue instead of sending them.
Only after all fragments belonging to the datagram pass the forwarding engine, all fragments queued in the VRB are sent.
The performance of FF in those experiments is comparable to the HWR results we observed in our evaluation above.
The number of link layer retransmissions also goes down to a comparable level. We consider this a strong indication of a consistent code base.

\section{Discussion}\label{sec:discussion}
In our testbed experiments, we were not able to reproduce the results for FF that are based on simulations as presented in \cite{tmw-6ff-19}. One striking difference between the two settings is our faster, lightly coordinated CSMA/CA MAC layer, which is by no means uncommon.
Corresponding problems have been already  hinted at in \cite{draft-ietf-6lo-minimal-fragment}, and are now substantiated.

We did not expect to see such disappointing results as revealed in this paper.
In our given scenario, FF becomes more of a hindrance than an improvement over HWR, even though our implementation optionally  falls back to HWR in case of fragment loss.
The only  advantage of FF we could clearly identify is its reduced RAM consumption.
Evaluating whether alternative approaches to fragment forwarding such as Selective Fragment Recovery \cite{draft-ietf-6lo-fragment-recovery} could help to mitigate these problems will be part of our future work.

Nonetheless, the stress on the device can only be reduced by a more elaborate MAC protocol.
In such attempts, however, care needs to be taken with the configuration of the experiment parameters:
Preliminary experiments with an existing MAC protocol in RIOT \cite{zs-gtammpi-17} led to problems such as frequent packet buffer overflows, after the packets stayed much longer  in the buffer queues of the MAC layer.

In the end, deployment scenarios and provider use cases should  decide whether fragment forwarding is applicable and if so on which MAC protocol.

\section{Conclusion and Outlook}\label{sec:conclusion}
In this paper, we evaluated direct fragment forwarding with 6LoWPAN in comparison to hop-wise reassembly using large real-world experiments.
We showed that with a  thin MAC layer, hop-wise reassembly can be the better choice to achieve proper reliability and latencies.
This contradicts previous results, but becomes clearer after careful analysis reveals that the medium is quickly exhausted by quicker fragment sending and retransmissions.

Further experiments are needed  
not only to evaluate more complex MAC layers and contrast with the results in~\cite{tmw-6ff-19}, but also to empirically relate FF to other fragment forwarding techniques including the selective 6LoWPAN fragment recovery protocol~\cite{draft-ietf-6lo-fragment-recovery}.
A possible direction of further evaluation could also include end-to-end fragmentation such as performed by IP~\cite{RFC-8200}.

\section*{A Note on Reproducibility}
We explicitly support reproducible research \cite{acmrep,swgsc-terrc-17}. Our experiments have been conducted in an open testbed. The source code of our implementations (including scripts to setup the experiments, RIOT measurement apps \etc) will be available on Github at \url{https://github.com/5G-I3/IEEE-LCN-2019}.

\section*{Acknowledgments}
We would like to thank Jakob Pfender and the anonymous reviewers for their
feedback on this paper.
This  work  was  supported  in  parts  by  the  German  Federal Ministry of Education and Research within the 5G project \emph{I3} and the VIP+ project \emph{RAPstore}.

\balance
\bibliographystyle{IEEEtran}
\bibliography{main,ids,own,manet,rfcs}

\begin{thebibliography}{10}
\providecommand{\url}[1]{#1}
\csname url@samestyle\endcsname
\providecommand{\newblock}{\relax}
\providecommand{\bibinfo}[2]{#2}
\providecommand{\BIBentrySTDinterwordspacing}{\spaceskip=0pt\relax}
\providecommand{\BIBentryALTinterwordstretchfactor}{4}
\providecommand{\BIBentryALTinterwordspacing}{\spaceskip=\fontdimen2\font plus
\BIBentryALTinterwordstretchfactor\fontdimen3\font minus
  \fontdimen4\font\relax}
\providecommand{\BIBforeignlanguage}[2]{{%
\expandafter\ifx\csname l@#1\endcsname\relax
\typeout{** WARNING: IEEEtran.bst: No hyphenation pattern has been}%
\typeout{** loaded for the language `#1'. Using the pattern for}%
\typeout{** the default language instead.}%
\else
\language=\csname l@#1\endcsname
\fi
#2}}
\providecommand{\BIBdecl}{\relax}
\BIBdecl

\bibitem{draft-ietf-6lo-plc}
J.~Hou, B.~Liu, Y.-G. Hong, X.~Tang, and C.~Perkins, ``{Transmission of IPv6
  Packets over PLC Networks},'' IETF, Internet-Draft -- work in progress~00,
  February 2019.

\bibitem{RFC-8200}
S.~Deering and R.~Hinden, ``{Internet Protocol, Version 6 (IPv6)
  Specification},'' IETF, RFC 8200, July 2017.

\bibitem{IEEE-802.15.4-15}
{IEEE 802.15 Working Group}, ``{IEEE Standard for Local and metropolitan area
  networks---Part 15.4: Low-Rate Wireless Personal Area Networks (LR-WPANs)},''
  IEEE, New York, NY, USA, Tech. Rep. IEEE Std 802.15.4\texttrademark--2015,
  April 2016.

\bibitem{RFC-4944}
G.~Montenegro, N.~Kushalnagar, J.~Hui, and D.~Culler, ``{Transmission of IPv6
  Packets over IEEE 802.15.4 Networks},'' IETF, RFC 4944, September 2007.

\bibitem{sb-6wei-09}
Z.~Shelby and C.~Bormann, \emph{6LoWPAN: The Wireless Embedded Internet},
  1st~ed.\hskip 1em plus 0.5em minus 0.4em\relax Wiley Publishing, 2009.

\bibitem{draft-ietf-6lo-minimal-fragment}
T.~Watteyne, C.~Bormann, and P.~Thubert, ``{6LoWPAN Fragment Forwarding},''
  IETF, Internet-Draft -- work in progress~03, July 2019.

\bibitem{RFC-7228}
C.~Bormann, M.~Ersue, and A.~Keranen, ``{Terminology for Constrained-Node
  Networks},'' IETF, RFC 7228, May 2014.

\bibitem{kkt-siotsp-14}
S.~L. Keo, S.~S. Komar, and H.~Tschofenig, ``{Securing the Internet of Things:
  A Standardization Perspective},'' \emph{IEEE Internet of Things Journal},
  vol.~1, no.~3, pp. 265--275, May 2014.

\bibitem{RFC-6282}
J.~Hui and P.~Thubert, ``{Compression Format for IPv6 Datagrams over IEEE
  802.15.4-Based Networks},'' IETF, RFC 6282, September 2011.

\bibitem{RFC-7400}
C.~Bormann, ``{6LoWPAN-GHC: Generic Header Compression for IPv6 over Low-Power
  Wireless Personal Area Networks (6LoWPANs)},'' IETF, RFC 7400, November 2014.

\bibitem{pttm-rpfri-18}
G.~Z. Papadopoulos, P.~Thubert, S.~Tsakalidis, and N.~Montavont, ``{RFC 4944:
  Per-hop Fragmentation and Reassembly Issues},'' in \emph{Proc. of the 2018
  IEEE Conference on Standards for Communications and Networking (CSCN)}.\hskip
  1em plus 0.5em minus 0.4em\relax IEEE, October 2018.

\bibitem{draft-ietf-lwig-6lowpan-virtual-reassembly}
C.~Bormann and T.~Watteyne, ``{Virtual reassembly buffers in 6LoWPAN},'' IETF,
  Internet-Draft -- work in progress~01, March 2019.

\bibitem{w-fsf6id-17}
A.~Weigel, ``{Forwarding strategies for 6LoWPAN-fragmented IPv6 datagrams},''
  Ph.D. dissertation, Technische Universit{\"a}t Hamburg-Harburg, 2017.

\bibitem{wrtt-roft6-13}
A.~Weigel, M.~Ringwelski, V.~Turau, and A.~Timm-Giel, ``{Route-over forwarding
  techniques in a 6LoWPAN},'' in \emph{International Conference on Mobile
  Networks and Management}.\hskip 1em plus 0.5em minus 0.4em\relax Springer,
  2013, pp. 122--135.

\bibitem{draft-ietf-6lo-fragment-recovery}
P.~Thubert, ``{6LoWPAN Selective Fragment Recovery},'' IETF, Internet-Draft --
  work in progress~05, July 2019.

\bibitem{cicrky-rmr6-09}
A.~H. Chowdhury, M.~Ikram, H.-S. Cha, H.~Redwan, S.~Shams, K.-H. Kim, and S.-W.
  Yoo, ``{Route-over vs Mesh-under Routing in 6LoWPAN},'' in \emph{Proc. of the
  2009 International Conference on Wireless Communications and Mobile Computing
  (IWCMC): Connecting the World Wirelessly}.\hskip 1em plus 0.5em minus
  0.4em\relax ACM, June 2009, pp. 1208--1212.

\bibitem{tmw-6ff-19}
Y.~Tanaka, P.~Minet, and T.~Watteyne, ``{6LoWPAN Fragment Forwarding},''
  \emph{IEEE Communications Standards Magazine}, vol.~3, no.~1, pp. 35--39,
  July 2019.

\bibitem{egvw-s6n-18}
M.~Esteban, D.~Glenn, M.~Vucinic, and T.~Watteyne, ``{Simulating 6TiSCH
  Networks},'' \emph{Wiley Internet Technology Letters}, vol.~30, no.~3, 2018.

\bibitem{anahi-6refr-18}
S.~A. Awwad, N.~K. Noordin, B.~M. Ali, F.~Hashim, and N.~H.~A. Ismail,
  ``{6LoWPAN Route-Over with End-to-End Fragmentation and Reassembly Using
  Cross-Layer Adaptive Backoff Exponent},'' \emph{Wireless Personal
  Communications}, vol.~98, no.~1, pp. 1029--1053, Jan. 2018.

\bibitem{bhgws-rotoi-13}
E.~Baccelli, O.~Hahm, M.~G{\"u}nes, M.~W{\"a}hlisch, and T.~C. Schmidt, ``{RIOT
  OS: Towards an OS for the Internet of Things},'' in \emph{Proc. of the 32nd
  IEEE INFOCOM. Poster}.\hskip 1em plus 0.5em minus 0.4em\relax Piscataway, NJ,
  USA: IEEE Press, 2013, pp. 79--80.

\bibitem{bghkl-rosos-18}
\BIBentryALTinterwordspacing
E.~Baccelli, C.~G{\"u}ndogan, O.~Hahm, P.~Kietzmann, M.~Lenders, H.~Petersen,
  K.~Schleiser, T.~C. Schmidt, and M.~W{\"a}hlisch, ``{RIOT: an Open Source
  Operating System for Low-end Embedded Devices in the IoT},'' \emph{IEEE
  Internet of Things Journal}, vol.~5, no.~6, pp. 4428--4440, December 2018.
  [Online]. Available: \url{http://dx.doi.org/10.1109/JIOT.2018.2815038}
\BIBentrySTDinterwordspacing

\bibitem{swgsc-terrc-17}
Q.~Scheitle, M.~W{\"a}hlisch, O.~Gasser, T.~C. Schmidt, and G.~Carle,
  ``{Towards an Ecosystem for Reproducible Research in Computer Networking},''
  in \emph{Proc. of ACM SIGCOMM Reproducibility Workshop}.\hskip 1em plus 0.5em
  minus 0.4em\relax New York, NY, USA: ACM, August 2017, pp. 5--8.

\bibitem{acmrep}
{ACM}, ``{Result and Artifact Review and Badging},''
  \url{http://acm.org/publications/policies/artifact-review-badging}, Jan.,
  2017.

\bibitem{plwhb-ownsr-15}
H.~Petersen, M.~Lenders, M.~W{\"a}hlisch, O.~Hahm, and E.~Baccelli, ``{Old Wine
  in New Skins? Revisiting the Software Architecture for IP Network Stacks on
  Constrained IoT Devices},'' in \emph{1st Int. Workshop on IoT Challenges in
  Mobile and Industrial Systems (IoT-Sys15)}.\hskip 1em plus 0.5em minus
  0.4em\relax Florence, Italy: ACM, May 2015.

\bibitem{lkhpg-cwemr-18}
\BIBentryALTinterwordspacing
M.~Lenders, P.~Kietzmann, O.~Hahm, H.~Petersen, C.~G{\"u}ndogan, E.~Baccelli,
  K.~Schleiser, T.~C. Schmidt, and M.~W{\"a}hlisch, ``{Connecting the World of
  Embedded Mobiles: The RIOT Approach to Ubiquitous Networking for the Internet
  of Things},'' Open Archive: arXiv.org, Technical Report arXiv:1801.02833,
  January 2018. [Online]. Available: \url{https://arxiv.org/abs/1801.02833}
\BIBentrySTDinterwordspacing

\bibitem{microchip:at86rf231}
\emph{{Low Power 2.4 GHz Transceiver for ZigBee, IEEE 802.15.4, 6LoWPAN, RF4CE,
  SP100, WirelessHART, and ISM Applications (AT86RF231)}}, Microchip, September
  2009, {Rev.8111C}.

\bibitem{microchip:at86rf233}
\emph{{Low Power, 2.4GHz Transceiver for ZigBee, RF4CE, IEEE 802.15.4, 6LoWPAN,
  and ISM Applications (AT86RF233)}}, Microchip, July 2014, {Rev. 8315E}.

\bibitem{zs-gtammpi-17}
S.~{Zhuo} and Y.~{Song}, ``{GoMacH: A Traffic Adaptive Multi-channel MAC
  Protocol for IoT},'' in \emph{Proc. of the 42nd IEEE Conference on Local
  Computer Networks (LCN)}.\hskip 1em plus 0.5em minus 0.4em\relax Piscataway,
  NJ, USA: IEEE Press, October 2017, pp. 489--497.

\end{thebibliography}

\appendix[Compile Time Parameters in RIOT]

\begin{table}[h]
    \centering
    \caption{Changed compile time parameters in RIOT.}
    \label{tab:impl:params}
    \begin{tabular}{>{\hspace{-.8em}}r>{\hspace{-.4em}}>{\hspace{-.4em}}r<{\hspace{-.6em}}}
        \toprule
        Compile-time Configuration Parameter & Value \\
        \midrule
        \scriptsize\texttt{GNRC\_NETIF\_PKTQ\_POOL\_SIZE} & \scriptsize $64$ \\
        \scriptsize\texttt{GNRC\_SIXLOWPAN\_FRAG\_RBUF\_SIZE} & \scriptsize $1$ (src.) / $16$ (sink) \\
        \scriptsize\texttt{GNRC\_SIXLOWPAN\_FRAG\_RBUF\_TIMEOUT\_US} & \scriptsize $10000000$ \\
        \scriptsize\texttt{GNRC\_SIXLOWPAN\_FRAG\_RBUF\_AGGRESSIVE\_OVERRIDE} & \scriptsize $0$ \\
        \scriptsize\texttt{GNRC\_SIXLOWPAN\_MSG\_FRAG\_SIZE} & \scriptsize $64$ \\
        \bottomrule
    \end{tabular}
\end{table}

\end{document}